\documentstyle[11pt]{article}
\begin{document}

\centerline{\bf\Large Applicability of the Friedberg-Lee-Zhao
method}
\vspace{1cm}

\centerline{Gong-Bo Zhao$^{1}$, Yi-Bing Ding$^2$, Xue-Qian Li$^1$,
Guo-Li Wang$^1$ and Zhaorigetu$^3$}
\vspace{1cm}

1. Department of Physics, Nankai University, Tianjin, 300071,
China.

2. Department of Physics, Graduate School, The Chinese Academy of the Sciences,
Beijing 100039, China.

3. Department of Physics, Chifeng Normal College of Inner-Mongolia, 024000, China.

\vspace{1cm}

\begin{center}
\begin{minipage}{10cm}
{\small Friedberg, Lee and Zhao proposed a method for effectively
evaluating the eigenenergies and eigen wavefunctions of quantum
systems. In this work, we study several special cases to
investigate applicability of the method. Concretely, we calculate
the ground-state eigenenergy of the Hellmann potential as well as the Cornell
potential, and also
evaluate the energies of the systems where linear term is added
to the Coulomb and harmonic oscillator potentials as a
perturbation. The results obtained in this method have a
surprising agreement with the traditional method or the numerical
results. Since the results in this method have obvious analyticity
compared to that in other methods, and because of the simplicity for
calculations this method can
be applied to solving the Schr\"{o}dinger equation and provides
us better understanding of the physical essence of the concerned
systems. But meanwhile applications of the FLZ method are
restricted at present, especially for certain potential forms, but due to
its obvious advantages, it should be further developed.}
\end{minipage}
\end{center}

\vspace{2cm}

\baselineskip 22pt

\noindent{\bf I. Introduction}\\

Recently, Friedberg, Lee and Zhao (FLZ) have proposed a method
\cite{Frie} which is very powerful and useful for solving the
Schr\"{o}dinger equation, especially for the potentials which have
strong couplings.

In their works \cite{Fried,Zhao}, applicability and
characteristics of the method were discussed,  on both the
physical and mathematical aspects.

As well known, only very few potential forms can lead to analytic
solutions of the Schr\"{o}dinger equation, whereas for the
majority of potentials which are of phenomenological significance
in real physical world, one can only expect numerical results. It
reduces the ability for further analyzing the problem and getting
insight into the physics essence. Therefore an almost-analytic
solution would be very welcome for studying the physics behavior
of the system.

In quantum mechanics, the WKB approximation is an important method
to study the potential-barrier tunneling  and quantization
conditions for bound states. In fact, the WKB approximation is an
semi-classical approximation where one can write the wavefunction
$\psi$ as $\psi(x)=exp({i\over\hbar}\alpha(x))$ and expand
$\alpha(x)$ according to orders of $\hbar$. Generally, only the
first two terms are retained \cite{Gasi}. It is confirmed to be
useful for tunneling rate evaluation, but for the bound states, it
can only offer a rough estimation.

Comparing with the WKB approximation,   FLZ proposed that the
wavefunction can be written as
\begin{equation}
\label{exp}
\phi(q)=e^{-s(q)},
\end{equation}
where $q$ is a set of any coordinates. Obviously, the WKB
wavefunction $exp({i\over \hbar}\alpha(x))$ is an oscillating form
which corresponds to real particles, so that more applicable to
the tunneling case, whereas, the FLZ wavefunction is a damping form
and more suitable for dealing with bound states.

In this work, we would like to investigate the applicability of
the FLZ method in terms of a few potentials which have
significance for practical physics problems. We first, in terms of
the FLZ method, evaluate the eigenenergy of the ground state for
the Cornell potential \cite{Corn} and the
Hellmann potential \cite{Hel}, and secondly we study
applications of this method to the perturbation calculations.
Concretely, we consider a linear term as a perturbation to the
hamiltonians which contain either the Coulomb potential(the Cornell potential) or
harmonic oscillator potential, because the first one has
importance for spectra of heavy quarkonia and the second has a
rigorous analytic solution.

We also employ the Dalgarno-Lewis technique \cite{DL} to obtain the
corresponding perturbation results up to the third order (for the
Cornell potential). Comparing the results with that
obtained in the FLZ method, one can find that they coincide with
each other order by order. However, obviously the procedure of
derivation in the FLZ method is much simpler.

It is noticed that the FLZ method cannot be applied to the cases
for linear potential $a_1q$, or the $1/q^2$ potential
(not for a perturbation). We will briefly
discuss this issue in the last section and
also concern other restrictions to applications of the method.

This paper is organized as following, after this introduction, we
briefly introduce the concerned aspects of the FLZ method and in
Sec.III, we discuss the Hellmann potential case and the Cornell
potential case, then in Sec.IV,
we apply  this method to the
perturbation calculations. The last section would be devoted to
the conclusion and discussion.\\

\noindent{\bf II. Brief introduction to the FLZ method}\\

For a completeness and convenience of readers, we briefly
introduce the Friedberg-Lee-Zhao method in this section.

The Schr\"{o}dinger equation can be written as
\begin{equation}
\label{Sch}
H\phi(q)=E\phi(q),
\end{equation}
where the hamiltonian  is
$$H={-1\over 2}\nabla^2+V(q)\;\;\;\;{\rm and}\;\;\;\;\;
\nabla^2=\sum_{i=1}^n{\partial^2\over \partial q_i^2}$$
and here for simplicity we take the effective mass $\mu$ to be unity.
Friedberg, Lee and Zhao suggest to write the wavefunction of the
ground state in the form of eq.(\ref{exp}) and moreover, they set
\begin{equation}
V(q)=g^2v(q),
\end{equation}
where $g^2$ is a scale factor and corresponds to the strength of
the effective coupling, for example in the QED Coulomb case, it
is $Ze^2$ and for the QCD Coulomb case, it is $\alpha_{eff}$
\cite{Corn}. The advantage of pulling out the coupling parameter
will be shown in our later examples.

Choosing $g^2$ as an expansion parameter, one can expect from the
expression of eq.(\ref{exp}) that the larger $g^2$ is, the faster
the expansions of $s(q)$ and $E$ converge.
Thus one can have expansions as
\begin{eqnarray}
\label{expan}
E &=& g^lE_o+g^{l-i}E_1+g^{l-2i}E_2+..., \\
s(q) &=& g^ms_o+g^{m-j}s_1+g^{m-2j}s_2+...,
\end{eqnarray}
where $l,m,i.j$ are positive integers and to be determined
according to a rule \cite{Fried} that under limit of strong
interaction,
\begin{equation}
l={2k\over n+2}
\end{equation}
where $n$ is the power of the leading term in the potential, for
example, $n=2,\; l=k/2$ for the harmonic oscillator and $n=-1,\;
l=2k$ for the Coulomb potential. To avoid difficulties in the
WKB approximations such as the matching conditions at the
turning points, it is required that $E_0$ and $v$ do not
appear in one equation and other equations are solvable. Thus one
has
\begin{eqnarray}
l<k:\;\;\;\; && 2m=k,\;\;\;\; 2m-j=l,\;\;\;\; i=j,\\
l>k:\;\;\;\; && 2m=l,\;\;\;\; 2m-j=k,\;\;\;\; i=j,
\end{eqnarray}
where $k$ is any positive integers which can make $l$ to be
an integer.

For example, in the simplest case, $l=m=i=j=1$, one has
\begin{eqnarray}
\label{set}
(\nabla s_0)^2 &=& 2v, \nonumber\\
\nabla s_0\cdot\nabla s_1 &=& {1\over 2}\nabla^2s_0-E_0, \nonumber \\
\nabla s_0\cdot\nabla s_2 &=& {1\over 2}(\nabla^2s_1-(\nabla
s_1)^2)-E_1,\nonumber \\
&& .....
\end{eqnarray}
It is noted that in the first equation, $E_0$ does not appear, and
only $v$ exists, whereas in the second, only $E_0$, but  $v$ is
absent. So in this method, it is not needed to distinguish
between $E<V$ or $E>V$ cases.

From the first equation of (\ref{set}), one can obtain $\nabla
s_0$ and then substitutes it into the second equation. Here it is
supposed that the potential is a function of $r$. For the
S-states, $\nabla^2={d^2\over dr^2}+{2\over r}{d\over dr}$, thus
$${d\over dr}s_m|_{r\to 0}$$
must be zero to insure $(\nabla^2 s_{m+1})|_{r\to 0}$ finite,
where $m$ is the order index. With this condition, we  obtain
$E_0$ and $\nabla s_0$ simultaneously from the second equation.
Then we will continuously achieve $E_{m-1}$ and $\nabla s_m$ from
the m-th equation.

With the boundary condition $\phi(0)=1$ or $s(0)=0$, we also
obtain the wavefunction up to a normalization factor.

Later we will show that this method is very powerful for
evaluating the eigenenergy of the ground state.\\

\noindent{\bf III. Applications to the Hellmann potential and the
Cornell potential}\\

(1) The Hellmann potential is a superposition of a Coulomb piece
and a Yukawa piece \cite{Hel}
\begin{equation}
\label{Hel}
V(r)=g^2({-A\over r}+{B\over r}e^{-Cr}),
\end{equation}
which cannot be analytically solved and here we deliberately pull
out a coupling constant $g^2$. In physics, the Coulomb potential usually
originates from a single massless photon or gluon exchange, but the Yukawa
potential is due to a massive particle exchange and can be induced by a form factor
at the effective vertices. Such a potentail may play an important role in phenomenology.
Now let us evaluate the
ground energy in terms of the FLZ method. We have the expansions
for $s$ and $E$ similar to (\ref{expan}), and here $l=4,i=j=m=2$,
then the set of equations is recast as
\begin{eqnarray}
\label{s0}
(\nabla s_0)^2 &=& -2E_0, \nonumber\\
\nabla s_0\cdot\nabla s_1 &=& {1\over 2}\nabla^2s_0-{A\over
r}+{B\over r}e^{-Cr}-E_1, \nonumber \\
\nabla s_0\cdot\nabla s_2 &=& {-1\over 2}(\nabla s_1)^2-{1\over 2}
\nabla^2 s_1-E_2,\nonumber \\
\nabla s_0\cdot\nabla s_3 &=& -\nabla s_1\cdot\nabla s_2+{1\over
2}\nabla^2 s_2-E_3,\nonumber \\
&& .....
\end{eqnarray}
The general expression for the even order of $n>1$ is
\begin{equation}
\nabla s_0\cdot\nabla s_{2n}=-\sum_{m=1}^{n-1}\nabla
s_m\cdot\nabla s_{2n-m}-{1\over 2}(\nabla s_n)^2+{1\over
2}\nabla^2s_{2n-1}-E_{2n},
\end{equation}
and for odd order, it is
\begin{equation}
\nabla s_0\cdot\nabla s_{2n+1}=-\sum_{m=1}^{n}\nabla
s_m\cdot\nabla s_{2n+1-m}+{1\over
2}\nabla^2s_{2n}-E_{2n+1}.
\end{equation}
Thus we obtain
$$s_0=\sqrt{-2E_0}r,$$
where we only keep the the positive root. The second equation is
written as
\begin{equation}
\label{sing}
\sqrt{-2E_0}{ds_1\over dr}={1\over r}\sqrt{-2E_0}-{A\over
r}+{B\over r}e^{-Cr}-E_1={\sqrt{-2E_0}-A+Be^{-Cr}\over r}-E_1.
\end{equation}
It is required that the left side of the equation and $E_1$ must be
finite at $r=0$, thus a natural condition
$$\sqrt{-2E_0}-A+Be^{-Cr}\rightarrow 0,\;\;\;{\rm as}\;\;
r\rightarrow 0,$$
is enforced. One  obtains immediately
\begin{equation}
E_0={-1\over 2}(A-B)^2\;\;\;{\rm and}\;\;\; s_0=|A-B|r.
\end{equation}
As a matter of fact, the key point of the FLZ method is to obtain
the first-order eigenenergy  by removing the
singularity in the equation and it is similar to the secular
equation for obtaining eigenenergies in quantum mechanics.

Because all $\nabla^2s_1$ (i=1,2...) are finite, so if we expand
$$s_i={a_{-n}^{(i)}\over r^n}+{a_{-n+1}^{(i)}\over r^{n-1}}+...
+{a_{-1}^{(i)}\over r}+a_0^{(i)}+a_1^{(i)}r+a_2^{(i)}r^2+...,$$
all $a_{-m}^{(i)}$ and $a_1$ must vanish, thus
$${ds_m\over dr}\rightarrow 0,\;\;\; {\rm as}\;\; r\rightarrow
0, \;\; (m\geq 1).$$
But $s_0$ is an exception which is linearly proportional to $r$,
indeed, $\nabla^2 s_0$ is singular at $r\to 0$. It is exactly the
condition to determine the value of $s_0$ as done in
eq.(\ref{sing}).

As we go on with the same strategy we  obtain other $E_i's$ and
one can find that all superficial singularities in the expressions
are cancelled out automatically. We write the eigen-energy of the ground state as
\begin{equation}
E={-1\over 2}(A-B)^2g^4-BC g^2+{3BC^2\over 4(A-B)}+O({1\over
g^{2n}}), \;\;(n\geq 1).
\end{equation}
Obviously as the coupling $g^2$ is sufficiently large, the higher
orders can be safely neglected.
Moreover, one can notice that as $A=B$ it blows up and we will discuss
this issue later.

To make sense, we take a set of the parameters to testify the
results. We can numerically solve the Sch\"{o}dinger equation and
obtain the numbers. Of course, we suppose that the
numerical results are precise. A comparison of the numerical results with
that with the FLZ method is shown in Table 1.

\vspace{1cm}

\begin{center}
\begin{tabular}{|c|c|c|}
\hline
g & Numerical result & FLZ \\
\hline
1 & -1.0848 & -0.75\\
\hline
2 & -11.3687 & -11.25\\
\hline
3 & -48.805 & -48.750\\
\hline
5 & -336.757 & -336.750\\
\hline
\end{tabular}\end{center}

\begin{center}
\begin{minipage}{13cm}
{\small Table 1. The comparison of the numerical results with
 that in terms of the FLZ method where only
the first three terms in the expansion are taken.
The parameters are set as $A=2,
 B=C=1$.}
\end{minipage}
\end{center}

\vspace{0.3cm}

It is noted that as $g=3\sim 5$, the sum of the first three terms from the result in the
FLZ method are perfectly  consistent with the numerical results.

(2) The Cornell potential.

The hamiltonian is
\begin{equation}
\label{total}
H={-1\over 2}\nabla^2-{\alpha_{eff}\over r}+\kappa r,
\end{equation}
for using the FLZ method let us transform it into another
convenient form
\begin{equation}
\label{nonper}
H={-1\over 2}\nabla^2+g^2({-1\over r}+\kappa' r),
\end{equation}
where $g^2$ is pulled out. The set of equations is similar to that for the Hellmann
potentail as ${A\over r}-{B\over r}e^{-Cr}$ is replaced by ${-1\over r}+\kappa r$.
\begin{eqnarray}
\label{Corn}
(\nabla s_0)^2 &=& -2E_0, \nonumber\\
\nabla s_0\cdot\nabla s_1 &=& {1\over 2}\nabla^2s_0-{1\over
r}+\kappa' r-E_1, \nonumber \\
\nabla s_0\cdot\nabla s_2 &=& {-1\over 2}(\nabla s_1)^2-{1\over 2}
\nabla^2 s_1-E_2,\nonumber \\
\nabla s_0\cdot\nabla s_3 &=& -\nabla s_1\cdot\nabla s_2+{1\over
2}\nabla^2 s_2-E_3,\nonumber \\
&& .....
\end{eqnarray}
The following procedures are exactly the same as in last subsection. We obtain
\begin{equation}
\label{kappa}
E={-1\over 2}g^4+{3\over 2}\kappa'-{3\over 2}\kappa^{'2}g^{-4}+{27\over 4}
\kappa^{'3}g^{-8}
+....
\end{equation}
One can notice that in the hamiltonian (\ref{nonper}), we set the
mass $\mu=1$ for simplicity. When we put back the mass, the
Schr\"{o}dinger equation is
\begin{equation}
\label{per1}
[{-1\over 2\mu}\nabla^2-{g^2\over r}+\kappa r]|\psi>=E|\psi>,
\end{equation}
the equation can be re-written as
\begin{equation}
\label{per2}
[{-1\over 2}\nabla^2-{g^2\mu\over r}+\mu\kappa r]|\psi>=\mu E|\psi>.
\end{equation}
Therefore, the expansion parameter $g^2$ is replaced by $\mu g^2$.
Even though in the practical case, the effective coupling
$\alpha_{eff}$ is about 0.4$\sim $0.5 being small, for heavy
quakonia such as $\Upsilon$, the reduced mass $\mu\sim 2.5$ GeV,
and $\mu g^2$ is not too small, the method can give result which
is close to the numerical result. Our numerical calculation indeed
shows that for the $b\bar b$ quakonium $\Upsilon$, the FLZ method
works very well, but for the $c\bar c$ system $J/\psi$ where
$\mu\sim 0.75$ GeV, the method fails to give a reasonable solution.

Now let us turn to the perturbation cases.\\

\noindent{\bf IV. Perturbation in the FLZ method}\\

(1) First we discuss a very simple case where a linear term is attached
to the harmonic oscillator as a perturbation, the hamiltonian is
\begin{equation}
H={1\over 2}{d^2\over dx^2}+{1\over 2}g^2x^2+ax,
\end{equation}
for simplicity, here we still take the mass $\mu=1$. The accurate ground state energy
is $E={1\over 2}g-{1\over 2}{a^2\over g^2}$. Thus we would use this result to testify the
applications of  the FLZ method for dealing with a perturbation and in
this situation $a$ is the expansion parameter. According to the general
rule and eq.(\ref{set}), we have
\begin{eqnarray}
\label{harm}
&& (-{d\over dx}s_0)^2+x^2 = 0, \nonumber \\
&& {d^2\over dx^2}s_0-2{ds_0\over dx}{ds_1\over dx}=2E_0,\nonumber\\
&&{d^2s_1\over dx^2}-({ds_1\over dx})^2-2{ds_0\over dx}{ds_2\over dx}
+2ax=2E_1,\nonumber\\
&&{d^2s_2\over dx^2}-2{ds_0\over dx}{ds_3\over dx}-2{ds_1\over dx}{ds_2\over dx}
=2E_2,\nonumber\\
&& .....
\end{eqnarray}
From the first and the second equations,
$$2E_0=1-2x{ds_1\over dx},$$
setting $x=0$, we obtain $E_0={1\over 2}$ and ${ds_1\over dx}=0$. As we go on calculating,
we can achieve
$$E_1=0,\;\; E_2=0,\;\; E_3={-1\over 2}a^2,\;\;{\rm and}\;\;E_4=E_5=...=0.$$
$$s_0={1\over 2}x^2,\;\;s_1=0,\;\; s_2=ax,...$$
So finally,
$$s={1\over 2}gx^2+{1\over 2g}ax,\;\;\;i.e.\;\; \phi=Nexp({-1\over 2}gx^2-{1\over 2g}ax),$$
$$E={1\over 2}g-{1\over 2}{a^2\over g^2}.$$
This energy is precisely the same as the exact solution.\\

(2) The Coulomb potential with a linear term as the perturbation.\\

In this subsection, we treat the linear term $\kappa r$ as a perturbation to the Coulomb piece.
The hamiltonian is
\begin{equation}
\label{per}
H={-1\over 2}\nabla^2-{g^2\over r}+\kappa r.
\end{equation}
The difference between eq.(\ref{per}) with eq.(\ref{total}) is the position of the
coupling $g^2$. Thus we have a set of equations as
\begin{eqnarray}
\label{pert}
(\nabla s_0)^2 &=& -2E_0, \nonumber\\
\nabla s_0\cdot\nabla s_1 &=& {1\over 2}\nabla^2s_0-{1\over
r}-E_1, \nonumber \\
\nabla s_0\cdot\nabla s_2 &=& {-1\over 2}(\nabla s_1)^2-{1\over 2}
\nabla^2 s_1+\kappa r-E_2,\nonumber \\
\nabla s_0\cdot\nabla s_3 &=& -\nabla s_1\cdot\nabla s_2+{1\over
2}\nabla^2 s_2-E_3,\nonumber \\
&& .....
\end{eqnarray}
again, the difference of the set from the set (\ref{Corn}) is the position of
$\kappa r$ which appears in the third equation, but in the set (\ref{Corn}),
it exists in the second equation.
Solving the equations according the normal procedure, we have
$$E_0={1\over 2},\;E_1=E_2=0,\; E_3={3\over 2}\kappa,\; E_4=E_5=0,\;
E_6={-3\over 2}\kappa^{2},\; E_7=E_8=0,\; E_9={27\over 4}\kappa^{3},...$$
$$s_1=r,\; s_1=0,\;s_2={1\over 2}\kappa r^2,\; s_3=0,\;s_4={-1\over 6\kappa^{3}},\;
s_5={-1\over 2\kappa^{2}}r^2,\;...$$
Thus the energy of the ground state is
\begin{equation}
\label{kappa1}
E={-1\over 2}g^4+{3\over 2}\kappa g^{-2}-{3\over 2}\kappa^{2}g^{-8}+{27\over 4}\kappa^{3}
g^{-14}+O(\kappa^{4}g^{-20})+....
\end{equation}
If we set $\kappa=\kappa' g^2$, we recover the expression (\ref{kappa}). The first term
(\ref{kappa1}) is the energy caused by the Coulomb potential alone. It
is amazing to notice that in this subsection, we treat $\kappa r$ as a
perturbation to the Coulomb potential, while in section III.(2), we
deal with the Coulomb and linear pieces as a whole potential, then
the two results are the same to the order $O(\kappa^3g^{-14})$.

For a better comparison, let us calculate the contribution from the linear term order by order
in the traditional perturbation theory.
The first order perturbation result can be easily obtained in the traditional method as
\begin{equation}
E_1=<R_{1,0}|\kappa r|R_{1,0}>={3\kappa\over 2g^2},
\end{equation}
where $R_{1,0}=2g^3e^{-g^2r}$.

For evaluating the second order and even higher orders
of corrections to the eigen-energy, the calculation procedure
in the method given by most textbooks of
quantum mechanics is too complicated to carry out practical calculations.
It fails because in the traditional method, all energy states including the
continuous and discrete spectra must be summed over. Dalgarno and Lewis introduced a
technique  \cite{DL}, which allows us to calculate
the perturbation contributions without carrying out the tedious summation
and obtain  results at any given order. The technique, in fact, is based on the
hypervirial theorem \cite{Lucha} and all details about the technique can be found
in ref.\cite{DL}.

By this technique,
we can immediately obtain
the second and the third order corrections to the energies of
the ground state of the Coulomb potential as
$$E_2={-3\over 2}{\kappa^{2}\over g^8},\;\; E_3={27\over 4}{\kappa^{3}\over g^{14}},$$
and this is exactly that we derived in the FLZ method.
\\

\noindent{\bf V. Conclusion and discussion}\\

From Secs.III and IV. where we discussed solutions for the Hellmann potential and the Cornell
potential which has importance to the heavy quarkonium,  and the perturbative
cases in the FLZ method. First we find that for strong coupling $g$ at the
Hellmann potential the results obtained in terms of the FLZ method are very
close to the numerical results, in fact the larger $g$ is, the closer to the
real solution the FLZ result is. We can conclude that the FLZ method applies
perfectly for strong interactions. Moreover, this method allows us to have
analytic forms for both eigenenergies and wavefunctions, so it has obvious advantages
for physical analysis.

Then we use the FLZ method to calculate the energy corrections due to a linear
perturbation to the harmonic oscillator and find that the result is exactly the
same as the precise solution.

We employ the FLZ method to calculate the energy for the Cornell potential
and pretend that the coupling is strong. Then alternatively, we treat the linear
piece as a perturbation to the Coulomb piece. We calculate the correction to
very high orders in the FLZ method and  then we employ the
Dalgarno-Lewis technique to repeat the calculation, amazingly, we find that
all the three methods give the same results (precisely), even though they
start from completely different points. This amazing consistence confirms the
applicability of the FLZ method. Moreover, the calculations with the FLZ method
are much simpler than with others
and one can avoid the tedious and complicated integration
and summation.
The solution of both energy and wavefunction
have almost-analyticity.
In fact, from our pedagogical examples, one can believe that
the FLZ method can apply to much more complicated perturbation terms, such as
the L-S coupling, tensor potentials and the relativistic corrections etc.

On other side, at present, the FLZ method can only effectively apply to
deal with the ground state.
FLZ showed \cite{Frie,Fried} that for a special case the method
can be applied to evaluate the eigen-energy of the first excited
state, but not for general cases yet.
How to generalize this method to calculate excited states in
general is an interesting task and should be further investigated.

The second serious drawback of the method is that the method cannot apply
to the potential where only positive power terms exist and the
leading order is the linear one. The reason can be understood as the following.
If only the linear term exists,  for example, all the equations in the (\ref{set})
become trivial, so that one cannot determine $E_i$ from them and the method fails.
There could be some way to remedy this problem, and we will present some
possible ways to solve in our later work.

 It is also shown in our
expression for the Hellmann potential, the resultant energy is
related to $A-B$, which sometimes appears  at denominator. As
$A\rightarrow B$, the corresponding term diverges. Turning to the
original expression of the Hellmann potential, as $A=B$, it just
is the case of a linear term as the leading order. Because  there exists
an exact solution for the S-states in the case of the linear potential $\kappa r$
and it is the well-known Airy function, therefore the superficial
divergence is associated to the problem in the FLZ method. We will
further investigate this issue in our future work.

Moreover, it requires a strong coupling if one wants to have a more accurate
result for a certain potential. For example, in Sec.III (2),
for the Cornell potential, we pretend or assume that the coupling is strong,
i.e. $g$ is large, however, in reality, the coupling in the Cornell potential
is $\alpha_{eff}={4\alpha_s\over 3}\sim 0.3\; to\; 0.5$ and not large at all.
But as shown above, the real expansion parameter is $\mu g^2$
instead of $g^2$, so as long as $\mu$ is large enough, $\mu g^2$
is also large and the FLZ method applies. This is the case for the
$b\bar b$ quarkonium $\Upsilon$, but does not apply to the case of $J/\psi$.
Thus we, in general,
so far cannot expect to simply calculate the charmonium ground states in this method yet.

On other side, the method has obvious advantages as discussed above, it
can give solutions of both energy and wavefunction of the ground state
which may be the most important subject to study in phenomenology of particle physics,
and the solutions are almost-analytic, so that provide possibility for discussing physics
of the concerned problems. Moreover, the simplicity of this method for
dealing with complicated potential forms is remarkable. Therefore the Friedberg-Lee-Zhao
method is absolutely applicable and advantageous. Of course it is worth further studies.\\

\noindent{\bf Acknowledgment}\\

This work is partially supported by the National Natural Science Foundation of China.
We would like to thank Prof. W.Q Zhao for great help and enlightening discussions.\\

\end{document}